\documentclass[a4paper]{spie}  
\usepackage{subcaption}
\usepackage{float}
\usepackage{amsmath,amsfonts,amssymb}
\usepackage{graphicx}
\usepackage[colorlinks=true, allcolors=blue]{hyperref}

\newcommand{\aj}{\textit{Astronomical Journal}}

\title{Hard X-Ray Focal-Plane Compton Spectro-Polarimeter: Detector Development and Sensitivity Evaluation}

\author[a,b]{Abhay~Kumar}
\author[c]{Tanmoy Chattopadhyay}
\author[a]{Santosh V. Vadawale}
\author[a]{N. P. S. Mithun}

\affil[a]{Physical Research Laboratory, Astronomy \& Astrophysics Division, Ahmedabad, 380009, India}
\affil[b]{INAF-IAPS, Via del Fosso del Cavaliere 100, 00133 Rome, Italy}
\affil[c]{Kavli Institute of Particle Astrophysics and Cosmology, Stanford University 452 Lomita Mall, Stanford, CA 94305, USA}

\authorinfo{Further author information: (Send correspondence to Abhay Kumar)\\Abhay Kumar: E-mail: abhay.kumar@inaf.it, Telephone:  +39 06 4993 4098\\ Current affiliation: INAF-IAPS, Rome, Italy}

\begin{document} 
\maketitle

\begin{abstract}
The scientific potential of X-ray polarimetry has long been recognized, yet the challenges of measuring polarization have left it largely unexplored, mainly in the hard X-ray regime. With the advent of hard X-ray focusing optics, sensitive focal-plane Compton polarimeters are now feasible. An early example is CXPOL (Compton X-ray Polarimeter), developed at Physical Research Laboratory (PRL), India, which demonstrated 20–80 keV polarimetric capabilities using a plastic scatterer and a CsI(Tl) absorber array\cite{chattopadhyay15}. The CXPOL prototype demonstrated polarimetric capabilities in the 20–80 keV range, establishing a foundation for further development.

Building on this concept, we evaluate a hard X-ray spectro-polarimeter employing a position-sensitive plastic scatterer surrounded by position-sensitive absorber detectors. This geometry enables efficient reconstruction of Compton events and allows combined polarimetric and spectroscopic measurements via interaction positions and deposited energies in the detectors. We evaluate key performance parameters of the revised configuration of the second version of the CXPOL. Using Geant4 simulations, we assess key performance parameters, including modulation factor, polarimetric efficiency, and expected
sensitivity with modern hard X-ray optics. We also present the  characterization results of first prototype of a 100×20×5 mm$^3$ NaI(Tl) absorber read out on both ends by silicon photomultiplier (SiPM) array operating in coincidence, evaluating energy and position resolution and light-output variation along the detector. The coincidence readout also reduces SiPM background by an order of magnitude. The results demonstrate the strong potential of a position-sensitive Compton-
based focal-plane instrument for next-generation hard X-ray spectro-polarimetry.

\end{abstract}

\keywords{CXPOL, X-ray polarimetry, Geant4, Compton Spectro-polarimeter, Focal plane detector, Scintillators, Si photomultiplier}

\section{Introduction}           
\label{sect:intro}

The importance of measuring X-ray polarization from astrophysical sources has long been recognized, as such measurements probe into the behavior of matter and radiation under extreme magnetic and gravitational fields at the close vicinity of compact objects. However, limitations in the instrumental sensitivity contributed largely to the lack of progress in the field of polarimetry, resulting in no dedicated X-ray polarimetry missions in a long time, until recently, when NASA's Imaging X-ray Polarimetry Explorer (IXPE) \cite{weisskopf2021} was launched in 2022. IXPE since its launch, has revolutionized the field of observational X-ray astronomy with sensitive polarization measurements for a copious of X-ray sources in 2–8 keV energy band. This breakthrough relies on two key technologies: high-sensitivity focal plane detectors and grazing-incidence focusing optics. Extending sensitive polarimetric capabilities into the hard X-ray band (20–80 keV) requires a similar pairing of hard X-ray focusing optics (e.g., NuSTAR-like multilayer mirrors) with sensitive focal plane polarimeters.

Since hard X-ray emission originates in the immediate proximity of compact objects, sensitive polarimetry in this energy band can probe gravity and magnetic field close to a black hole or neutron star\cite{krawczynski11,chattopadhyay2021_review, soffitta10, soffitta2021polarized, jahoda2019x}. In particular, hard X-ray polarization measurements can provide critical insights into several astrophysical phenomena. These include the coronal geometry in Active Galactic Nuclei (AGN) and Black Hole Binary (BHB) systems, the magnetic field configurations and particle acceleration mechanisms in X-ray pulsars and magnetars, and the disk–jet interplay occurring near the base of jets in black hole systems.

Indian Space Research Organization (ISRO) has recommended 2-3 major space experiments in the coming future with focal plane hard X-ray polarimetry being one of them\footnote{\url{https://www.isro.gov.in/media_isro/pdf/Highlights/MSV2035_Astronomy_Astrophysics.pdf}}. In this context, PRL is developing a focal plane Compton polarimeter alongside a parallel development of a hard X-ray focusing optics\cite{tiwari2025development, tiwari2025enhanced, NeerajTiwari2025ExPA}. The conceptual design and initial performance of the first prototype instrument were reported previously in Chattopadhyay et al. 2013, 2014, 2015\cite{chattopadhyay13,chattopadhyay14_cxpol,chattopadhyay15}. This initial configuration utilized a central plastic scatterer read out by a photomultiplier tube (PMT), surrounded by a cylindrical array of sixteen CsI(Tl) scintillator absorbers coupled to individual silicon photomultipliers (SiPMs). While successfully demonstrating the proof of concept for the polarimetric geometry and readout architecture, evaluation of the initial prototype provided key insights and identified clear pathways to maximize the instrument sensitivity \cite{chattopadhyay15}.

Building upon the initial CXPOL framework, this work presents an upgraded focal-plane Compton spectro-polarimeter design featuring position-sensitive detectors and an improved absorber configuration. We describe the refined conceptual design of the instrument, detail key performance enhancements related to the first-generation polarimeter, and present experimental characterization results for the detector elements surrounding the scatterer. Finally, we summarize our findings and discuss the future plans for the development of the Compton polarimeter.

\section{Compton polarimeter working principle}

\begin{figure}[h!]
\centering
  \includegraphics[scale=0.35]{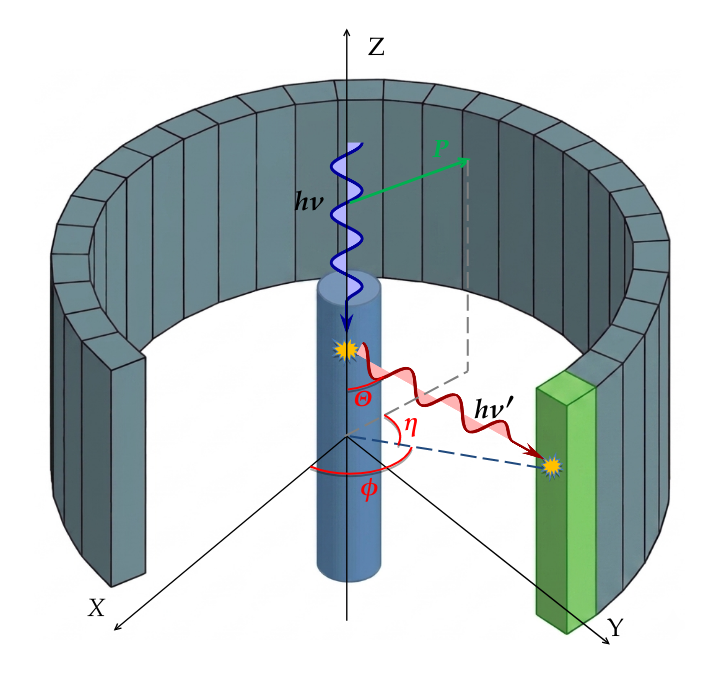}
\caption{Schematic of Compton scattering of a linearly polarized photon.}
\label{fig:compt_scatt}
\end{figure}

Compton scattering is the dominant process of photon - matter interaction in 20-80 keV. The Compton scattering is sensitive to the polarization of the incident photon. Polarisation induces a
preferential azimuthal angular direction of scattering (normal to the incident beam axis) as described by the Klein-Nishina cross section \cite{heitler54}:

$$\frac{\mathrm{d}\sigma}{\mathrm{d}\Omega}=\frac{r_{0}^{2}}2\left(\frac{E^{\prime}}E\right)^{2}\left(\frac E E+\frac{E^{\prime}}E-2\sin^{2}\theta\cos^{2}\eta\right)$$

where,
$${\frac{E^{\prime}}{E}}=\left[1+{\frac{E}{m_{\mathrm{e}}c^{2}}}(1-\cos\theta)\right]^{-1}$$

where $r_\circ$ is the classical electron radius, m$_{e}$c${^2}$ denotes the rest mass energy of an electron, E ($h\nu$) and E$^{'}$ ($h\nu^{'}$) are the incident and scattered  photon energy, respectively, $\theta$ is the Compton scattering angle, and $\eta$ forms the angle between the plane of scattered photon direction with the plane containing the polarization direction of the incident photon as shown in Figure \ref{fig:compt_scatt}.

The azimuthal distribution of scattering angles modulate as $\cos^{2}\eta$ and peaks in the direction orthogonal to the polarisation direction of the incident photons, implying linearly polarised photons are preferentially scattered orthogonal to their polarisation direction. The polarisation sensitivity of a polarimeter is defined in terms of modulation factor $\mu(\theta)$, which is the fraction of modulated signal corresponding to 100$\%$ polarised radiation. It is given by

$$\mu(\theta) = \frac{C_{\max}(\theta) - C_{\min}(\theta)}{C_{\max}(\theta) + C_{\min}(\theta)} = \frac{\left(\frac{d\sigma}{d\Omega}\right)_{\phi=\frac{\pi}{2}} - \left(\frac{d\sigma}{d\Omega}\right)_{\phi=0}}{\left(\frac{d\sigma}{d\Omega}\right)_{\phi=\frac{\pi}{2}} + \left(\frac{d\sigma}{d\Omega}\right)_{\phi=0}} = \frac{\sin^2\theta}{\frac{E}{E'} + \frac{E'}{E} - \sin^2\theta}$$

The modulation factor corresponding to a 100 $\%$ linearly polarized photon beam is called $\mu_{100}$. The polarisation angle and degree of polarisation are retrieved from the distribution of azimuthal scattering angle $\varphi$ by fitting the distribution with a cosine-squared function. The amplitude of the fitted distribution gives the modulation factor, and the phase $\pm$90 gives the polarisation angle. The polarimeter sensitivity is given by the Minimum Detectable Polarization (MDP)\cite{weisskopf10} at $99\%$ confidence level:

\begin{equation}
\text{MDP}_{99\%} = \frac{4.29}{\mu_{100} R} \sqrt{\frac{R + B}{T}}
\end{equation}

where $R$ is the source count rate, $B$ the background rate and $T$ the observing time.
The modulation factor $\mu$, instrument polarimetric efficiency $\epsilon$, and quality factor $Q$ are defined as

\begin{equation}
Q = \mu \sqrt{\epsilon}
\end{equation}

The polarimetric efficiency incorporates the tagging efficiency, defined as the probability of detecting an energy deposit in the scatterer, once the scattered photon have been detected in the absorber\cite{fabiani13}. 
The quality factor represents the sensitivity of the instrument based on its geometry only without taking into account the source brightness and background level.

\section{CXPOL V1: Compton X-ray polarimeter}\label{comp_spec_pol}

\begin{figure}[H]
\centering
  \includegraphics[scale=0.26]{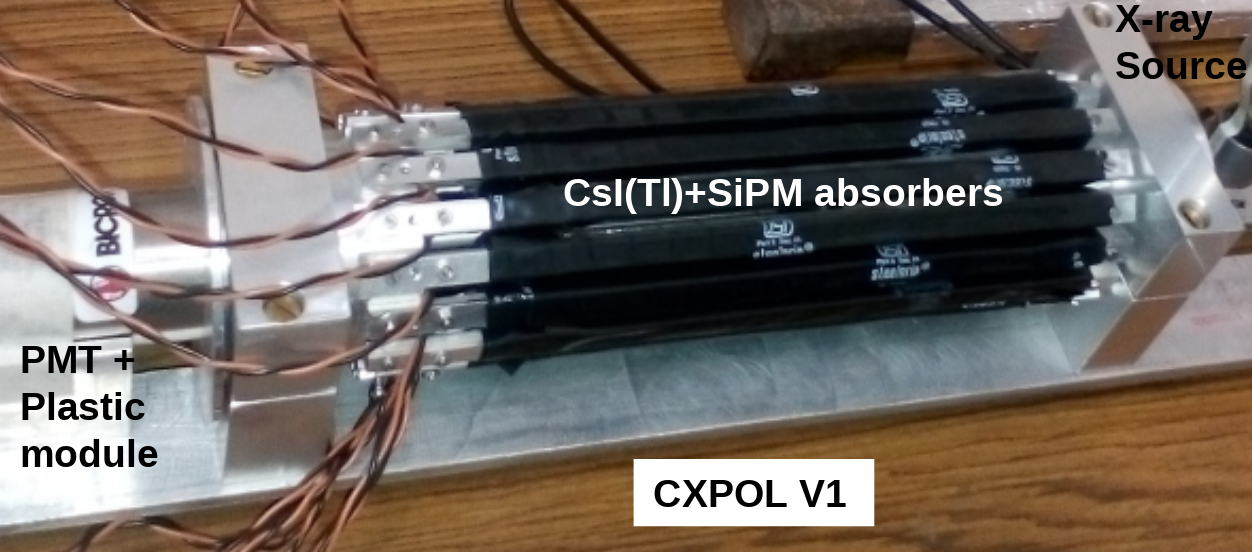}
  
\caption{Polarization experiment set up with the fully integrated configuration of the Compton polarimeter. The polarized source of radiation (shown in the figure) employs 90${^\circ}$ Compton scattering of the unpolarized photons from radioactive sources}
\label{fig:CXPOLV1}
\end{figure}

The Physical Research Laboratory (PRL), India, has been developing a compact focal-plane Compton X-ray polarimeter (CXPOL) coupled with NuSTAR type grazing incidence focusing optics, operating in the $20\text{--}80\text{ keV}$ energy range to establish technology readiness for prospective Indian space astronomy missions. Following the advent of hard X-ray optics for NuSTAR, the initial prototype (CXPOL v1, see Figure \ref{fig:CXPOLV1}) was developed as a proof-of-concept focal-plane polarimeter, with its design and initial performance reported in Chattopadhyay et al. 2013, 2014, 2015\cite{chattopadhyay13, chattopadhyay14_cxpol, chattopadhyay15}.

\begin{figure}[H]
\centering
  \includegraphics[scale=0.26,trim={2cm 0cm 2cm 0.cm},clip]{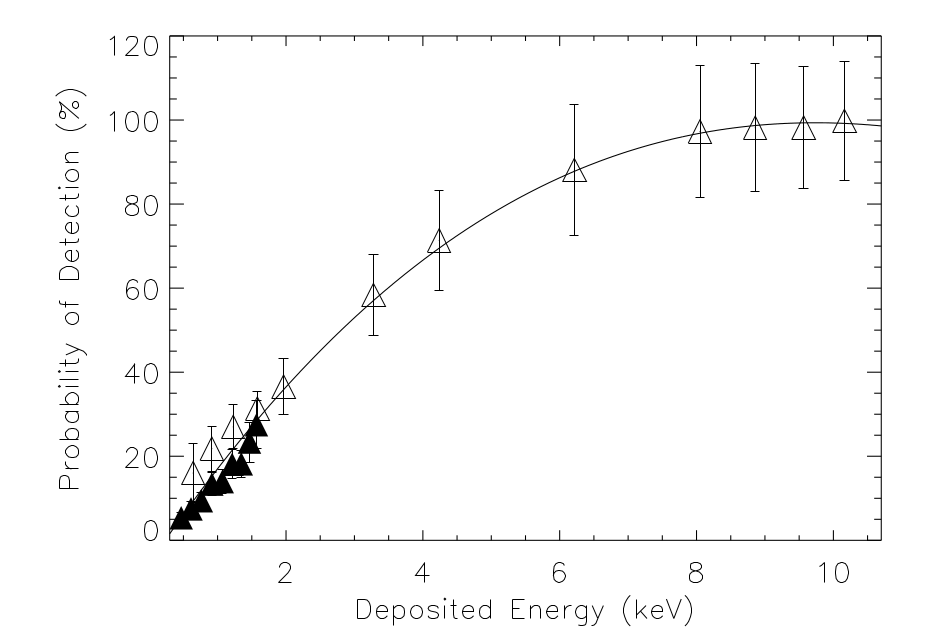}
  \includegraphics[scale=0.17,trim={0.3cm 0cm 1cm 0.cm},clip]{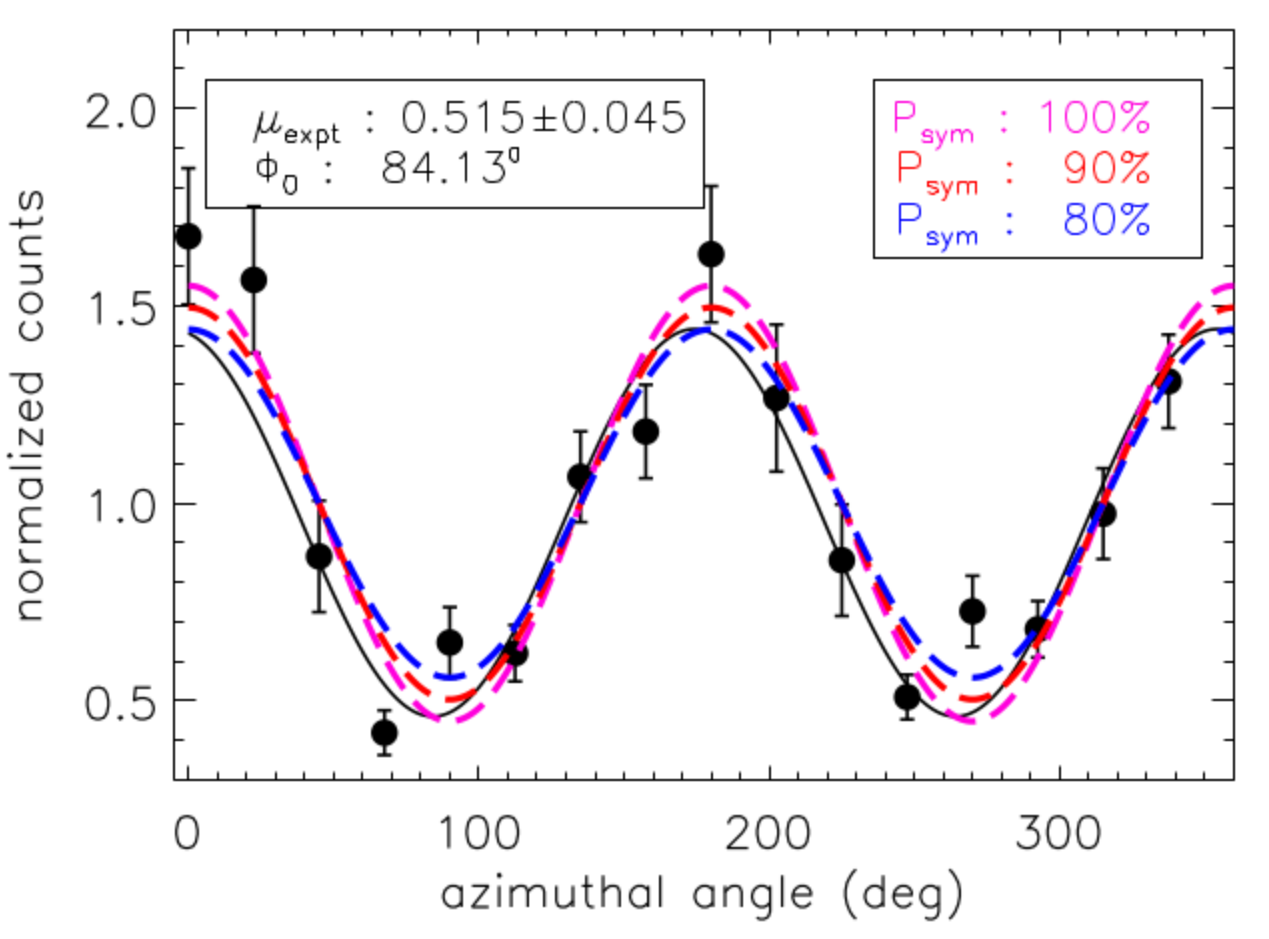}
\caption{\textbf{Left:} Detection probability as a function of deposited energy from 0.4 keV to 10 keV. Filled and open triangles correspond to 22.2 keV and 59.5 keV photons, respectively. These data points have been fitted with an empirical polynomial shown by the solid line. \textbf{Right:} Azimuthal angle distribution for partially polarized 20-50 keV continuum radiation for 90$^\circ$ polarization angles. The black solid line is the fit to the experimental data. The pink, red and blue dashed lines represent the modulation curves obtained from simulation for this setup for 100, 90 and 80$\%$ polarized beams respectively. This figure is taken from Chattopadhyay thesis, 2016, with permission}
\label{fig:cxpolv1_reslts}
\end{figure}

The conceptual geometry consists of a central, thin plastic scatterer surrounded by a circular array of high-$Z$scintillator absorbers designed to measure the azimuthal dependence of scattered X-ray photons. The central scatterer utilizes a cylindrical plastic scintillator (BC404; $5\text{ mm}$ diameter, $100\text{ mm}$ height) housed inside a $1\text{ mm}$ thick aluminum cylinder having a $0.5\text{ mm}$ aluminum entrance window. The scatterer unit, integrated with a photomultiplier tube (PMT; Hamamatsu R6095, bialkali photocathode with $\sim 25\%$ maximum quantum efficiency at $420\text{ nm}$), was procured from Saint-Gobain. Surrounding this central unit are sixteen $\text{CsI(Tl)}$ scintillator crystals (each $15\text{ cm}$ long with a $5\text{ mm} \times 5\text{ mm}$ cross-section). Each $\text{CsI(Tl)}$ crystal is wrapped with a reflective aluminum foil and enclosed in an aluminum casing open only on the face towards the scatterer and at the bottom end for readout. Each absorber is read out from one end by a single Silicon Photomultiplier (SiPM; KETEK PM3350, $3\text{ mm} \times 3\text{ mm}$ active area). In-house discrete preamplifiers, shaping amplifiers, and fast comparators were developed for energy extraction and timing trigger generation across all channels. Systematic characterization was conducted for both the plastic scatterer and individual $\text{CsI(Tl)}$-SiPM detector units. Performance evaluations of the central scatterer demonstrated $100\%$ detection efficiency for energy deposits above $\sim 7\text{ keV}$, below which the efficiency gradually degrades (see left panel of Figure \ref{fig:cxpolv1_reslts}).

Following comprehensive characterization of all the sixteen CsI absorbers, the fully integrated polarimeter was tested using both unpolarized and partially polarized X-ray beams. Trigger signals from the sixteen absorbers and the central scatterer were put in coincidence using FPGA based readout system, which applied $6~\mu\text{s}$ coincidence window to identify the genuine Compton events before transmitting data to a LabVIEW-based acquisition interface. Exposure to unpolarized radiation produced a spurious modulation amplitude consistent with zero, implying insignificant systematic effects which can arise due to non-uniformity in absorber sensitivity. Furthermore, tests were conducted with partially polarized radiation generated via $90^\circ$ primary scattering off an aluminium target using radioactive sources. The measured polarisation matched closely with the expected theoretical values  across multiple polarization orientation angles \cite{chattopadhyay15} (see Figure \ref{fig:cxpolv1_reslts}).

\section{CXPOL V2: Design modifications and optimisation }

The initial prototype demonstrated the proof-of-concept for both the polarimetric configuration and the associated readout electronics. However, it also revealed key areas for sensitivity optimization \cite{chattopadhyay15, kumar2026development}. Specifically, light collection in the CsI(Tl) absorbers was restricted to a few centimeters near the SiPM coupling face. Both the light collection efficiency and energy response degraded significantly beyond 4-5 cm from the readout end along the $15\text{ cm}$ length of the crystal. This degradation was primarily driven by the relatively slow $\sim1~\mu\text{s}$ primary decay time of $\text{CsI(Tl)}$, which dispersed photon arrival times and therefore reduced the signal-to-noise ratio (SNR).

\begin{figure}[H]
\centering
  \includegraphics[angle=-90, width=0.5\textwidth]{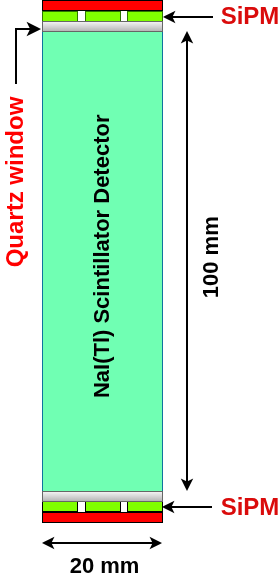}
   
\caption{ The schematic of the absorber detector (100$\times$20$\times$5 mm$^{3}$) with two end readout by the SiPMs. This figure is adapted from kumar et al\cite{kumar2026development}, with permission}
\label{fig:cxpolv2_absorber_sch}
\end{figure}

To address these limitations, several structural, optical, and sensor upgrades are implemented in the Version 2 (CXPOL V2) of the polarimeter particularly in the absorber units. First, the geometry of the absorber is re-optimized by reducing the length to $10\text{ cm}$ and increasing the cross-sectional width to $2\text{ cm}$, thereby expanding the effective light-collection area. Second, $\text{CsI(Tl)}$ is replaced with $\text{NaI(Tl)}$, a faster scintillator. The slow $\sim1~\mu\text{s}$ decay constant of $\text{CsI(Tl)}$ previously resulted in a lower instantaneous photon flux at the photodetector, yielding poor signal to noise. The faster decay time of $\text{NaI(Tl)}$ (230 ns) is expected to significantly improve the signal to noise ratio (SNR) and sensitivity along the length of the scintillator. Third, the KETEK SiPMs are replaced with the next-generation, low-noise SiPMs from Onsemi. Finally, the single-ended readout scheme is replaced with a dual-ended readout configuration (see Figure \ref{fig:cxpolv2_absorber_sch}). This dual-ended design not only maximizes light collection efficiency along the crystal length but also enables 1D interaction-position sensing along the length of the detector for simultaneous spectroscopy using Compton Kinematics.

To optimize the length of the central scatterer, we performed Monte Carlo simulations using the Geant4 simulation toolkit \cite{agostinelli03}. These simulations evaluate the modulation factor for a $100\%$ linearly polarized beam and the instrument efficiency, to optimise the quality factor across various scatterer lengths. For each energy starting from $20\text{ keV}$ and particular polarization angle, $10^6$ incident photons are simulated. 
To accurately model the focal-plane beam, the Geant4 General Particle Source (GPS) is configured as a circular spot with a 5 mm diameter, matching the physical cross-section of the scatterer. The output for each photon detected in the plastic scatterer and the $\text{NaI(Tl)}$ absorbers are stored as an event list for post-processing in Python. In the analysis pipeline, energy deposition thresholds are set to $1\text{ keV}$ for the plastic scatterer and $19\text{ keV}$ for the $\text{NaI(Tl)}$ absorbers. For the data analysis, single coincident events defined as an energy deposition exceeding the threshold in the scatterer alongside a single hit in the absorber are filtered to extract the azimuthal scattering angle. It is estimated using the interaction position information in
the scatterer and absorber. The estimation of polarisation degree and angle can be carried out using different techniques proposed in the literature\cite{DiMarco2022,sk13,kislat2015}. This assessment is carried out by exploring the technique of Strokes parameters estimation based on an unbinned event-by-event approach\cite{DiMarco2022,rankin2022,kumar2025cubesat}. 

\begin{figure}[H]
\centering

    \includegraphics[width=0.57\textwidth]{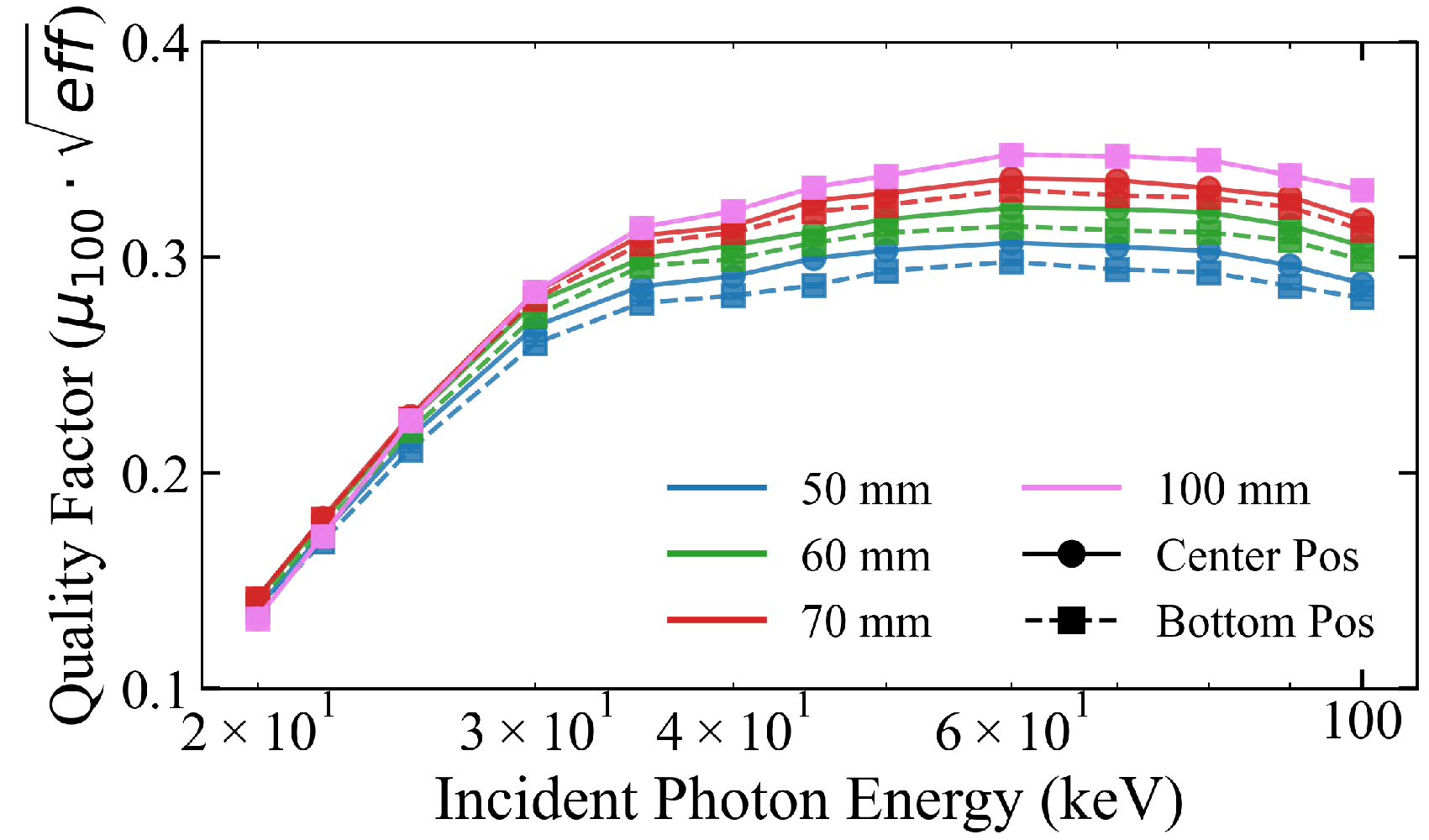}
\caption{Quality factor as a function of energy for different scatterer length and position}
\label{fig:cxpolv2_scatterer_sim}
\end{figure}

While a 10 cm plastic scatterer showed maximum polarimetric sensitivity, a 7 cm long BC404 plastic scatterer centered along the $z$-axis is expected to have better light collection, timing response, and position sensitivity under dual-end readout of the plastic (see Figure \ref{fig:cxpolv2_scatterer_sim}). This ensures sensitivity along the entire length of the scatterer while enabling simultaneous spectral reconstruction via Compton kinematics, provided the interaction positions in both the scatterer and absorbers are known.

\begin{figure}[H]
\centering
  \includegraphics[width=0.3\textwidth]{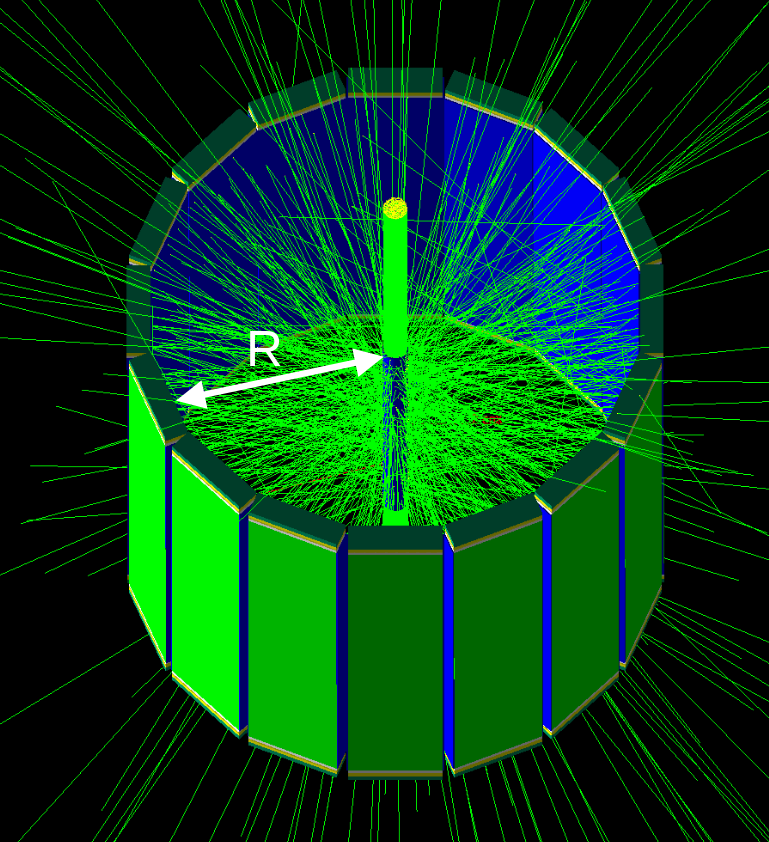}
  \includegraphics[width=0.55\textwidth]{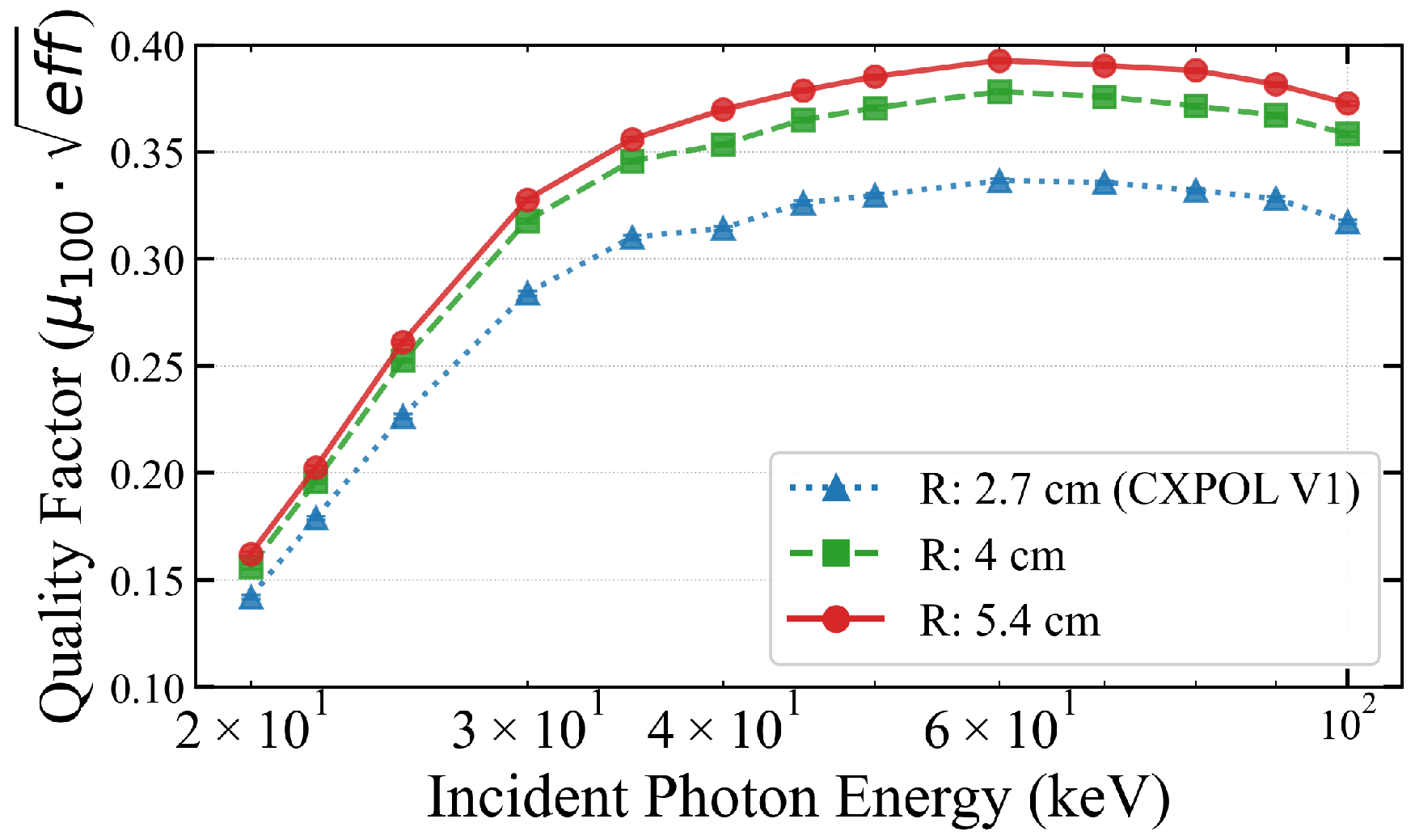}
\caption{\textbf{Left:} Conceptual design of the Compton spectro-polarimeter (CXPOL V2). 16 NaI(Tl) (5 mm $\times$ 20 mm $\times$ 100 mm) scintillators are shown in green, SiPM array in dark green and plastic scatterer (5 mm diameter and 70 mm length) in blue color at the center. The gaps between the scintillators are also seen, these are filled with 0.2 mm Al. \textbf{Right:} Quality factor as a function of energy for different scatterer and absorber distance (R)}
\label{fig:cxpolv2_design_opt}
\end{figure}


We further optimized the instrument geometry to determine the required scatterer to absorber radial distance ($R$) and the total number of absorber modules. The radial spacing introduces a performance trade-off: increasing $R$ improves the azimuthal angular resolution (more angular bins) and the modulation factor ($\mu_{100}$), but reduces the geometric detection efficiency. Conversely, a smaller radius degrades the modulation factor and limits the number of possible azimuthal bins. Using Geant4 Monte Carlo simulations across three candidate configurations, we identified an optimal radial distance of $R = 5.4\text{ cm}$ to maximize overall polarimetric sensitivity without making the number of channels indiscriminately large. Enclosing the central scatterer at this distance requires a total of 16 $\text{NaI(Tl)}$ absorber modules. The final configuration of the current prototype, therefore, consists of a $7\text{ cm}$ long plastic scatterer surrounded by sixteen dual-ended readout $\text{NaI(Tl)}$ absorbers (see Figure \ref{fig:cxpolv2_design_opt}).

The final updated design of the polarimeter with respect to the first version of the detector prototype is compared in the Table \ref{table:cxpolv1_v2_comp}.
\begin{table}[H]
    \centering
    \caption{Details of the scatterer and absorber along with scattering geometry.}
    \label{vt_thrsld}
    \begin{tabular}{p{6cm}|p{4cm}|p{4cm}}
        \hline
        \textbf{Parameters} & \textbf{CXPOL V1} & \textbf{CXPOL V2} \\
        \hline
        \multicolumn{3}{l}{\textbf{Scatterer}} \\
        \hspace{1em} Shape & Cylindrical & Cylindrical \\ 
        \hspace{1em} material & plastic & plastic \\ 
        \hspace{1em} Height & 100 mm & 70 mm \\
        \hspace{1em} Diameter & 5 mm & 5 mm \\
        \hspace{1em} Cover & 0.2 mm thick Aluminium & 0.2 mm thick Aluminium \\
      
        \multicolumn{3}{l}{\textbf{Absorber}} \\
        \hspace{1em} Shape & Cylindrical array & Cylindrical array \\
        \hspace{1em} material & 16 CsI(Tl) scintillator & 16 NaI(Tl) scintillator \\
        \hspace{1em} Dimension & 5 mm $\times$ 5 mm $\times$ 150 mm & 5 mm $\times$ 20 mm $\times$ 100 mm \\
        \hspace{1em} Dead space between absorbers & 0.2 mm Al & 0.2 mm Al \\
        \hspace{1em} Distance: scatterer center to front & 27.1 mm & 50.4 mm \\
        \hspace{1em} Al thickness (scatterer to absorber) & 0.2 mm & 0.2 mm \\
        \hspace{1em} Front cover & 0.2 mm epoxy & 0.2 mm epoxy \\
        \hspace{1em} Remaining sides & 0.2 mm Al & 0.2 mm Al \\
        \hline
    \end{tabular}
    \label{table:cxpolv1_v2_comp}
\end{table}

\section{Position sensitive absorber development}

\begin{figure}[H]
\centering
  \includegraphics[scale=0.42]{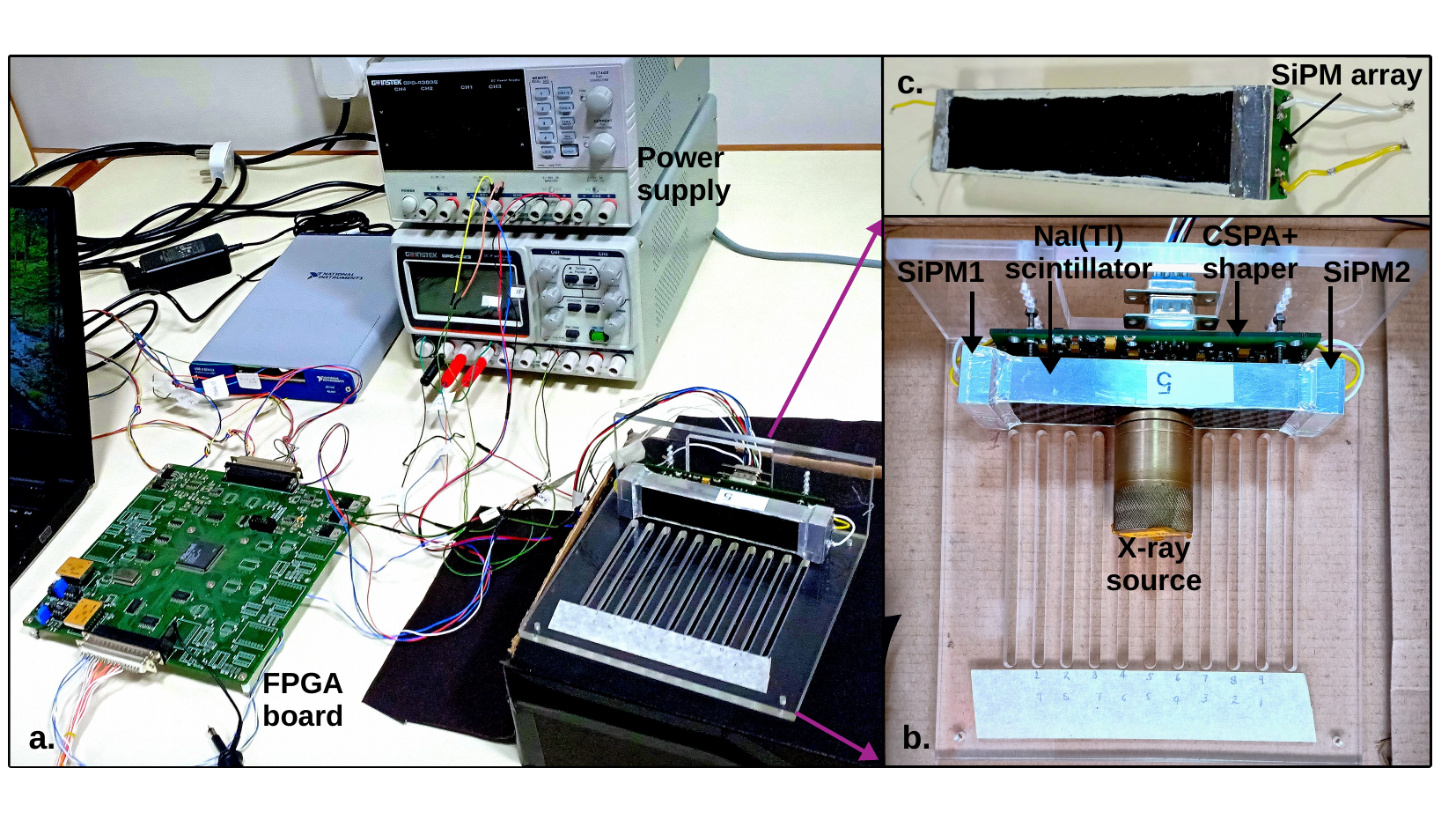}
\caption{\textbf{Panel a:} Experimental setup for the position measurement consisting of detector setup, front-end, back-end readout electronics, and power supply. 
		\textbf{Panel b:} NaI(Tl) scinitllator is packaged in the aluminium cover with SiPM array at both ends, and CSPA+Shaper is shown. \textbf{Panel c:} The bare scintillator readout by an array of 3 SiPMs at both ends is shown. This figure is taken from kumar et al. 2026\cite{kumar2026development}, with permission}
\label{fig:cxpolv2_absorber}
\end{figure}

The absorber detector module comprises a $\text{NaI(Tl)} $scintillator read out by SiPM arrays at both ends \cite{kumar2026development}. Due to the highly hygroscopic nature of$\text{NaI(Tl)}$, custom sealed detector modules are procured from Advatech, UK. The $\text{NaI(Tl)}$ scintillator is wrapped in polytetrafluoroethylene (PTFE a.k.a Teflon) tape on all four longitudinal faces, leaving only the two optical coupling faces open. The pre-wrapped crystal is housed inside an aluminum enclosure. This aluminum casing encloses three sides of the module, while the front entrance face oriented toward the central scatterer is covered by a $0.2\text{ mm}$ thick carbon window to minimize low-energy X-ray attenuation. Each optical end of the scintillator is coupled to an array of three SiPMs through quartz window and connected in series, providing a single summed signal output per end. Reflection losses at the optical interfaces (scintillator-quartz and quartz-SiPM window) are minimized using $0.5\text{ mm}$ thick layers of optically clear silicone gel. Finally, to eliminate stray ambient light, both readout ends are enclosed with aluminum caps (see Figure \ref{fig:cxpolv2_absorber}).

\begin{figure}[H]
    \centering
    \begin{subfigure}[b]{\textwidth}
        \centering
        \includegraphics[width=0.43\textwidth]{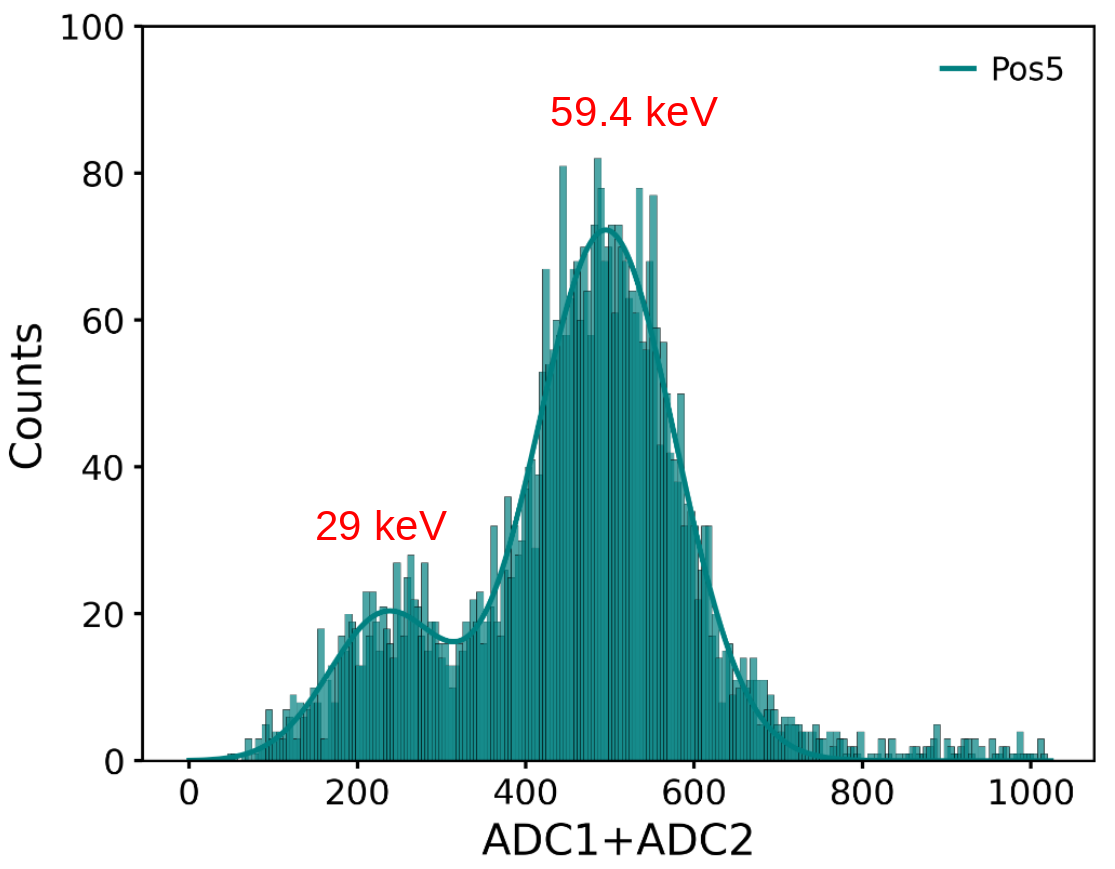}
    \end{subfigure}
    \hfill 
    \begin{subfigure}[b]{0.45\textwidth}
        \centering
        \includegraphics[width=\textwidth]{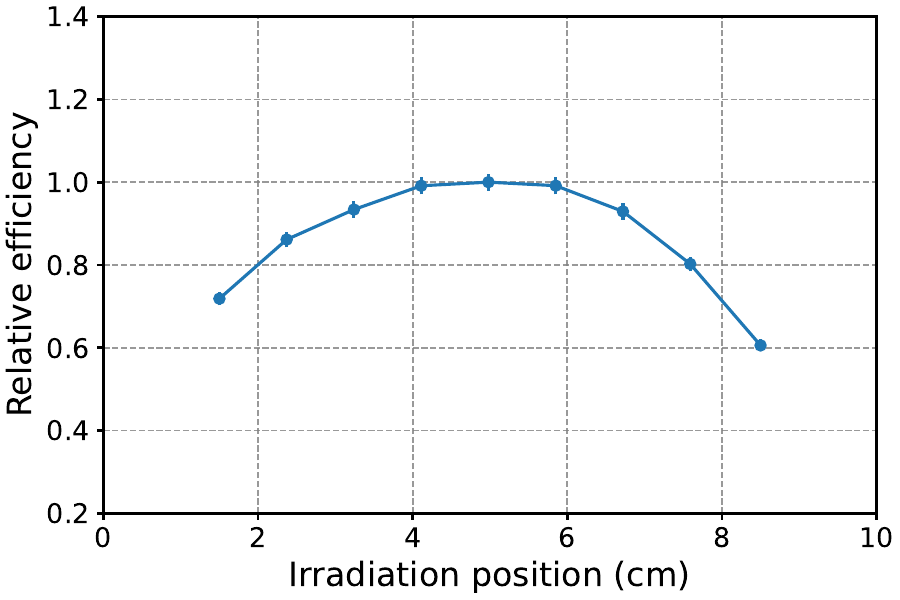}
    \end{subfigure}
    \hfill 
    \begin{subfigure}[b]{0.45\textwidth}
        \centering
        \includegraphics[width=\textwidth]{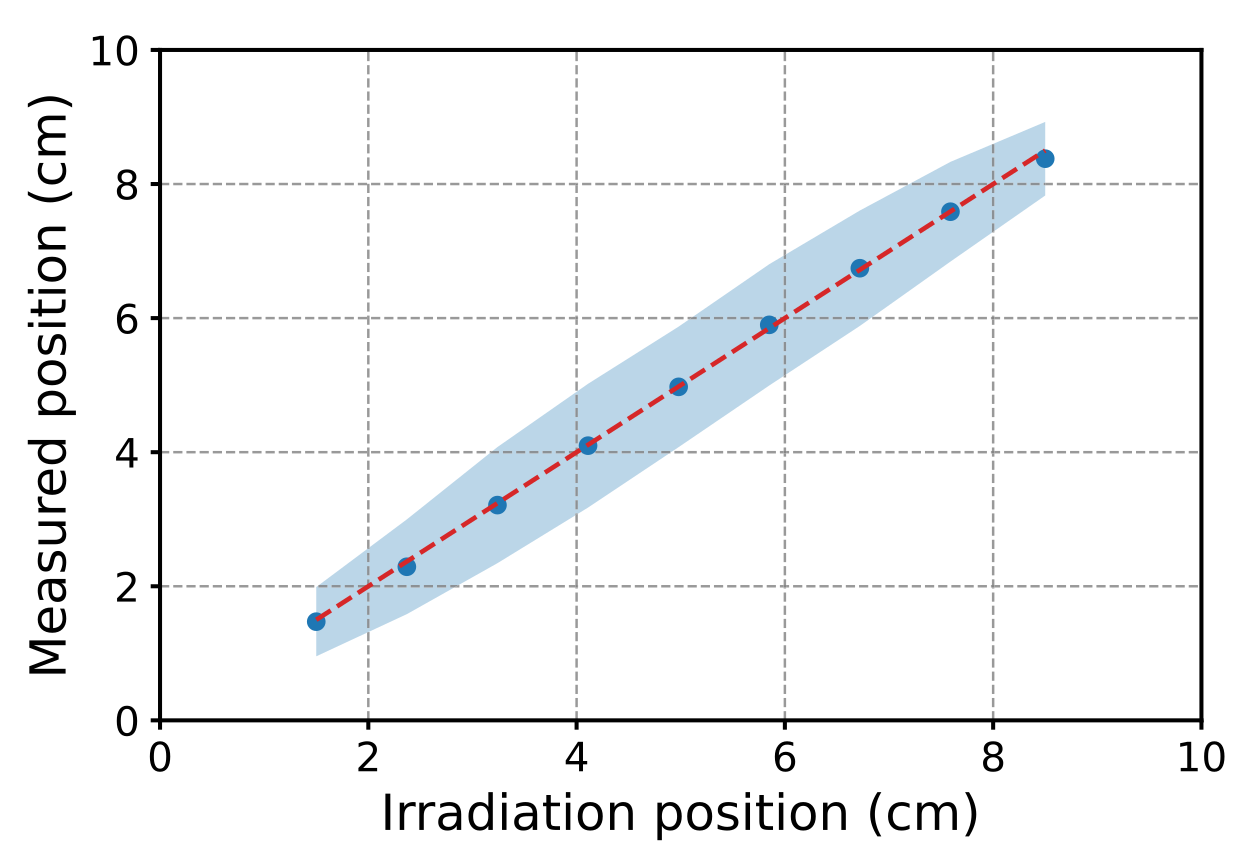}
    \end{subfigure}
    
    \caption{\textbf{Top: }The $^{241}\text{Am}$ spectra at the center of the detector obtained by summing the signal (ADC1+ADC2) at the two end of the detector. \textbf{Left:} Relative efficiency as a function of irradiation position. \textbf{Right:} Measured vs irradiation position along the length of the detector. Red dashed line represents the expected trend between irradiation position and measured position (blue markers). The shaded blue region represents the position resolution at each point. This figure is taken from kumar et al. 2026\cite{kumar2026development}, with permission}
    \label{fig:cxpolv2_absorber_reslts}
\end{figure}

Characterization of the dual-readout $\text{NaI(Tl)}$ absorbers is performed using an $^{241}\text{Am}$ radioactive source ($59.5\text{ keV}$) to evaluate detector performance along the length of the detector. Summing the signals from both readout ends ($\text{ADC}1 + \text{ADC}2$) provide the spectra. We can detect the $59.5\text{ keV}$ photopeak along with the iodine $K$-shell escape peak across the entire length of the absorber. The overall light collection efficiency peaked at the center of the crystal bar where optical attenuation path lengths to both photodetectors are equal and decreased symmetrically toward both ends. The interaction position along the length is determined using the ratio of the light outputs from the two SiPM arrays ($[\text{ADC}1 / \text{ADC}2]$). The average position resolution of $\sim 1.5\text{ cm}$ and energy resolution of 35$\%$ are achieved at 59.4 keV with the first prototype measurements (see Figure \ref{fig:cxpolv2_absorber_reslts}). Further improvements are expected through refinements in optical coupling, enhanced light-tight housing design and ASIC based readout electronics.

The dark count rate in SiPMs predominantly caused by thermally generated charge carriers that undergo avalanche multiplication and mimic genuine scintillation signals constitutes the primary contribution to total instrumental background. However, implementing the coincidence condition between the two opposing readout faces of a single absorber module significantly suppresses these uncorrelated thermal dark counts. This dual-ended coincidence filtering reduces the raw background rate by an order of magnitude compared to a single-ended trigger scheme\cite{kumar2026development}. This will in turn improve the polarimetric sensitivity of the instrument.

\section{CXPOL V2: Expected sensitivity}
\begin{figure}[H]
\centering
\includegraphics[scale=0.6]{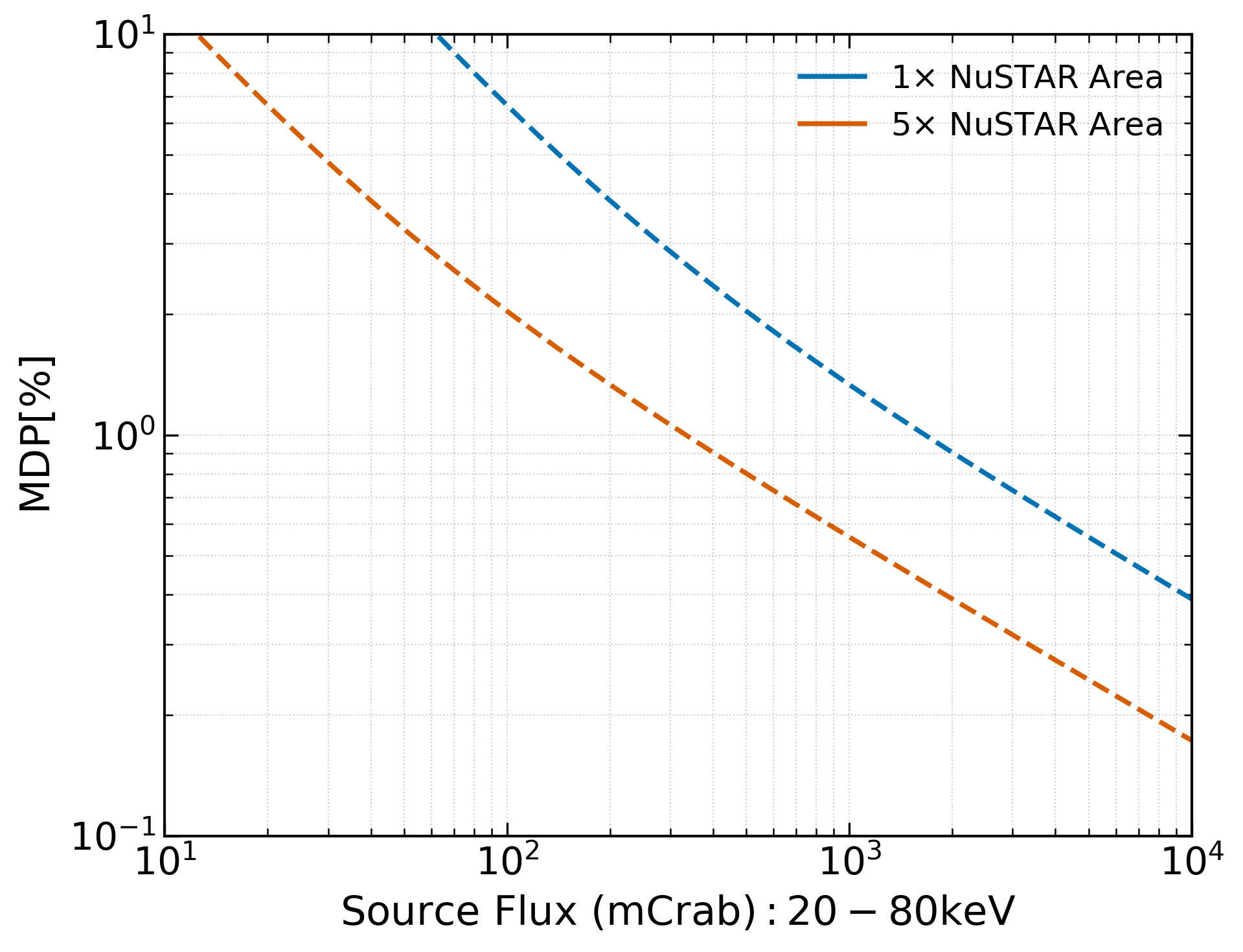}
\caption{Simulated polarimetric sensitivity of CXPOL V2 as a function of source intensity for a 100 ks of exposure time. The blue line represents result for single NuSTAR mirror area. The orange line refer to 5 times larger collecting area. The sensitivity calculations include the experimentally measured detection sensitivity of the scatterer and absorber.}
\label{fig:cxpolv2_senstivity}
\end{figure}

By incorporating the energy dependent detection efficiencies of both the central scatterer and surrounding absorbers, we calculated the polarimetric sensitivity of the optimized instrument for various source flux levels using Geant-4 simulations. Sensitivity is defined in terms of the Minimum Detectable Polarization at the $99\%$ confidence level ($\text{MDP}_{99}$). For a $100\text{ mCrab}$ point source observed for $100\text{ ks}$ using a NuSTAR like effective area ($\sim 500\text{ cm}^2$ at $30\text{ keV}$), the polarimeter achieves an $\text{MDP}_{99}$ of $\sim 7\%$ in 20-80 keV energy range assuming the integrated background count rate of 0.5 counts.s$^{-1}$. Scaling the collecting aperture area by a factor of five ($5\times$NuSTAR effective area) improves the sensitivity  to  $\sim 2\%$ for the same exposure time (see Figure \ref{fig:cxpolv2_senstivity}). These estimates represent conservative baseline performance metrics. Further sensitivity optimizations will be refined through the experimental calibrations of scatterer and absorber detector systems.

\section{Summary and future plans}
A dedicated hard X-ray polarimetry mission has been recently recommended by the Indian Space Research Organization (ISRO) for a possible future space mission from India. The Physical Research laboratory (PRL) has been working on the development of an engineering model of a focal plane Compton polarimeter, dubbed as CXPOL. Building upon the initial version of the CXPOL prototype, we have optimized the design and configuration of the focal-plane hard X-ray Compton polarimeter. Here we summarize the key findings $-$

\begin{itemize}
    \item The major change is in the absorber, where the $15\text{ cm}$ long $\text{CsI(Tl)}$ absorber is replaced with $10\text{ cm}$ long $\text{NaI(Tl)}$ scintillators coupled to dual-ended SiPM readouts. This configuration effectively resolves prior poor light-collection efficiency and provides sensitivity throughout the length of the detector.
    \item By implementing dual-ended readout, position sensitivity with resolution of $\approx 1.5\text{ cm}$ along the absorber length is achieved. It also suppresses SiPM thermal dark noise by an order of magnitude in  dual-ended coincidence readout. 
    \item Scatterer size is also optimized to $7\text{ cm}$ to have better light collection and position sensitivity in dual-ended readout which we will consider to implement in the future.
\end{itemize}

Based on these improvements, simulations indicate that for a $100\text{ mCrab}$ X-ray source in 20-80 keV and a $100\text{ ks}$ integration time for a single NuSTAR-like mirror module ($\sim 500\text{ cm}^2$ effective area) with a conservative background rate of $0.5\text{ counts s}^{-1}$, the polarimeter achieves a Minimum Detectable Polarization ($\text{MDP}_{99})$ of  $\sim 7\%$. Increasing the optics collecting area by a factor of five improves the sensitivity to $\text{MDP}_{99}$ to $\sim 2\%$.

Initial characterisation of the prototype NaI(Tl) absorber detector
also brings up a few aspects of possible further improvement in its design and readout system which will be implemented in the next version of the polarimeter.

\begin{itemize}
   
 \item The NaI(Tl) absorber detector is tested at 60 keV and it provides sensitivity throughout the length in coincident mode. However, it needs to be extended down to 20 keV for the Compton polarimeter with a targeted 20 keV polarimetric threshold. It is planned to test the detector systems down to 20 keV and improving upon the optical coupling of the scintillator with the SiPMs for better light collection.

 \item We aim to replace the discrete electronics with relatively faster Application Specific Integrated Circuits (ASICs) based readout electronics, which will result in shorter integration of SiPM dark counts. Such an upgrade should significantly improve the effective background rejection through tighter coincidence windows.  

 \item We also plan to test scintillators with high light yield, faster decay constants like $\text{CeBr}_3$ and GAGG for improved light collection which will in turn improve the energy resolution and low-energy threshold limits.

 \item Segmented or position-sensitive scatterers will also be explored for simultaneous spectroscopy using Compton kinematics.

\end{itemize}

These aspects are planned to be incorporated in the next version of the CXPOL type focal plane Compton polarimeter. 

\bibliographystyle{spiebib} 

\end{document}